# Pyrite–Bismuth Telluride Heterojunction for Hybrid Electromagnetic-to-Thermoelectric Energy Harvesting

Karthik R[1], Yiwen Zheng[2], Caique Campos de Oliveira[3], Punathil Raman Sreeram[1*], Pedro Alves da Silva Autreto[3*], Aniruddh Vashisth[2*], Chandra Sekhar Tiwary[1*]

[1]*Department of Metallurgical and Materials Engineering, Indian Institute of Technology Kharagpur, West Bengal, India*

[2]*Department of Mechanical Engineering, University of Washington, Seattle, WA, USA*

[3]*Center for Natural and Human Sciences (CCNH), Federal University of ABC (UFABC), Santo André 09210-580, Brazil*

*corresponding email: sreerampunam@metal.iitkgp.ac.in (PRS), pedroautreto@hotmail.com (PASA), vashisth@uw.edu (AV), chandra.tiwary@metal.iitkgp.ac.in (CST)*

**Abstract**

The rapid proliferation of wireless networks and connected devices has led to pervasive electromagnetic (EM) energy dissipation into the environment an underutilized resource for energy harvesting. Here, we demonstrate a pyrite ($FeS_2$)–bismuth telluride ($Bi_2Te_3$) heterojunction that enables hybrid electromagnetic-to-thermoelectric energy conversion. Fabricated via a simple cold-press compaction of powders, the heterojunction forms a Schottky interface at $FeS_2$, facilitating efficient RF absorption and localized heating. This heat is harvested by $Bi_2Te_3$ through thermoelectric conversion. Under 35 MHz RF irradiation at 1 W input power, the device achieved a local temperature rise of 46 °C and a thermal gradient of 5.5 K across the $Bi_2Te_3$, resulting in a peak power density of ~13 mW/cm². Molecular dynamics (MD) simulations and density functional theory (DFT) calculations further elucidate the heat transport behaviour and interfacial thermoelectric performance. This work introduces a new class of heterostructures for RF-responsive energy harvesting, offering a scalable route toward self-powered IoT and wireless sensing systems.

**Keywords**- Electromagnetic Energy harvesting, Thermoelectric, Iron sulfide, Bismuth telluride, Density functional theory, Molecular Dynamics.

**Introduction**

Our living environment provides several natural energy sources, including solar, wind, heat, and vibrations, and artificial energy sources such as electromagnetic waves. These energy sources have been effectively utilized in the development of various energy harvesting technologies, including Piezoelectric [1], triboelectric [2], thermoelectric [3–5], Photovoltaic [6,7], Magnetic field energy harvesting [8,9] and Radiofrequency (RF) energy harvesting [10]. While piezoelectric and triboelectric energy harvesting rely on moving components, RF and thermoelectric energy harvesting operate without such mechanical dependencies, making them more durable. Additionally, with the rapid expansion of wireless technologies such as 5G, IoT devices, and wireless power transmission, there has been a significant rise in electromagnetic energy density in the environment [11]. While most studies on RF energy harvesting have focused on designing rectenna-based systems (integrating antennas and rectifiers to convert RF signals into DC power directly), relatively little attention has been given to RF to thermal energy conversion[12]. Electromagnetic waves, particularly in the radio frequency range, have low energy and long wavelengths, making them challenging to capture efficiently. Conventional rectennas require precise frequency tuning and often operate efficiently only within narrow frequency bands [13]. An alternative and less-explored approach involves converting RF energy into localized thermal energy, which can then be harvested using thermoelectric materials to generate electricity.

Thermoelectric materials are widely recognized for their ability to convert thermal energy into electricity, making them ideal for energy harvesting through waste heat recycling and renewable energy generation. Various material systems, including metal oxides [14], metal tellurides [15,] and high entropy alloys [16] have demonstrated excellent thermoelectric properties. An ideal thermoelectric material is characterized by a high figure of merit (ZT), which depends on the optimal balance between thermal conductivity and electrical conductivity [17,18]. Significant progress has been made in enhancing the thermoelectric and functional properties of these materials. However, their interaction with electromagnetic waves remains relatively unexplored. The high thermal conductivity and electrical conductivity of thermoelectric materials typically result in lower electromagnetic energy absorption. While microwave-to-thermoelectric energy conversion has been demonstrated [19], harvesting lower-energy electromagnetic waves, such as radio waves, remains a significant challenge due to their longer wavelengths and lower photon energy. Efficient absorption of these waves requires semiconductors with extremely low band gaps to facilitate charge excitation. Integrating such low band gap semiconductors with thermoelectric materials offers a promising pathway to not only convert ambient RF energy into usable electrical power but also to mitigate electromagnetic interference (EMI). This synergistic approach addresses dual challenges, energy sustainability and electromagnetic pollution, making it highly relevant for next-generation wireless and IoT eco systems. To address the challenges outlined above, this work presents the development of a Radio Frequency-Thermoelectric (RFTE) heterojunction comprising $FeS_2$-$Bi_2Te_3$ constructed using straightforward techniques such as cold pressing. **Figure 1** shows the fabrication of the $FeS_2$-$Bi_2Te_3$ heterojunction and the photograph of the resultant heterojunction pellet. Pyrite ($FeS_2$), an earth-abundant semiconducting material with a low band gap of 0.8-0.9 eV, serves as a key component as it can absorb RF energy in high-frequency regions [20,21]. The Ti/$FeS_2$ heterojunction functions as a Schottky junction, enabling the detection of radio waves (RF) and their conversion into thermal energy. Subsequently, $Bi_2Te_3$, a well-known prominent thermoelectric material, then efficiently converts this thermal energy into electrical energy [22]. The Schottky heterojunction generated a maximum temperature of 45.5 °C at 35 MHz RF, and the overall RF-TE composite delivered a peak power output density of 13 mW/cm$^2$. The integration of sustainable materials with TE material for developing RF-TE devices positions FeS$_2$-Bi$_2$Te$_3$ as a promising candidate for the conversion of electromagnetic energy into electricity.

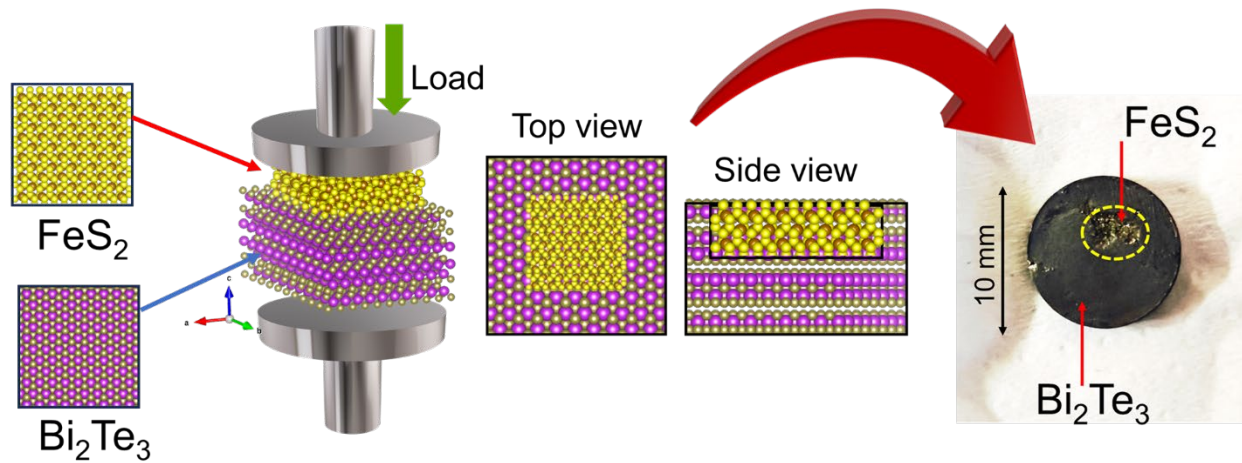

**Figure 1**- Schematic representation of the fabrication of the FeS$_2$–Bi$_2$Te$_3$ heterojunction and the corresponding digital photograph of the resulting pellet.

## Results and discussions

The structural and crystalline nature of the pellet was analyzed using X-ray diffraction (XRD), as shown in **Figure 2a**. The XRD pattern (Figure 2a) reveals peaks corresponding to Bi$_2$Te$_3$, indexed to hexagonal Bi$_2$Te$_3$ (ICSD:74348), and FeS$_2$, indexed to cubic FeS$_2$ (ICSD:53529). A more detailed morphological analysis of the FeS$_2$-Bi$_2$Te$_3$ heterojunction was performed using scanning electron microscopy (SEM) and energy-dispersive spectroscopy (EDS) studies as illustrated in **Figure 2(b-f).** The cross-sectional SEM image distinctly highlights the heterojunction between Bi$_2$Te$_3$ and FeS$_2$, while the EDS elemental maps show the elemental distribution across the heterojunction. The vibrational nature of the heterojunction was analysed using Raman spectroscopic studies as shown in **Figure 2g**. The peaks at 105.2 cm$^{-1}$, 122.6 cm$^{-1}$, 140 cm$^{-1}$ correspond to $E^2_g$, $A_{1u}$, $A^2_{1g}$ modes in Bi$_2$Te$_3$ [23,24] and the peaks at 342 cm$^{-1}$, 376 cm$^{-1}$, 428 cm$^{-1}$ correspond to $A_g$, $E_g$, and $T_g$ modes in FeS$_2$ [25–27]. Further, the chemical oxidation states of the pellet were analyzed using X-ray photoelectron spectroscopy (XPS).

**Figure 2h** shows the XPS survey spectrum of the sample, indicating the core levels of Bi 4p, Te 3d, Fe 2s, Fe2p, and S 2p. **Figure 2i** shows the Fe 2p binding energy spectrum with distinct peaks at 708.4 eV, 711.6 eV, 721.2 eV, and 730.2 eV. The intense peaks at 708.4 eV and 721.2 eV correspond to Fe $^{2+}$ states, confirming the formation of Fe (II)-S [28,29]. The peak at 711.6 eV is attributed to $Fe^{3+}$-O surface states, arising from surface oxidation. The binding energy plots of S 2p and Bi 4f are presented together in **Figure 2j** due to their proximity in binding energy. For S 2p, the peak at 162 eV corresponds to $S^{2-}$ core level, indicating the formation of $FeS_2$, while a less intense peak at 168.8 eV is attributed to $SO_4^{2-}$, resulting from surface oxidation. In the case of $Bi_2Te_3$, the Bi 4f spectrum displays peaks at 157.2 eV and 162.5 eV, corresponding to the $Bi^{3+}$ oxidation state, which is characteristic of $Bi_2Te_3$. In the Te 3d binding energy plot (**Figure 2k**), the peaks at 572.2 eV and 582.5 eV indicate the presence of $Te^{2-}$ oxidation states, while a minor peak at 574 eV is attributed to $TeO_2$, resulting from surface oxidation [30,31]. The above studies confirm the structural and chemical stability of $FeS_2$-$Bi_2Te_3$.

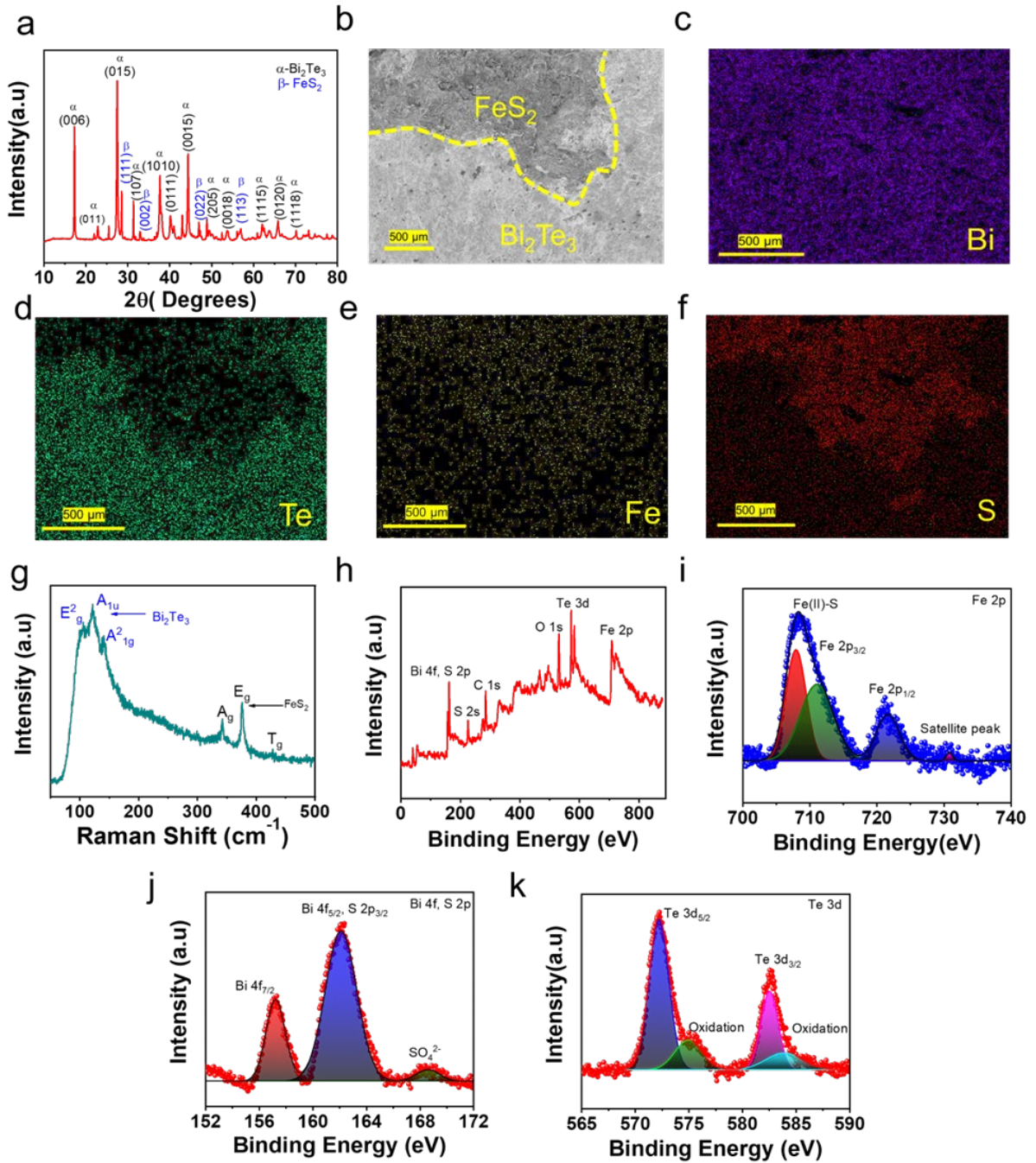

**Figure 2** – (a) Photograph of FeS$_2$-Bi$_2$Te$_3$ pellet, (b) XRD pattern of FeS$_2$-Bi$_2$Te$_3$ pellet, (c) FESEM image of FeS$_2$-Bi$_2$Te$_3$ heterojunction, (d-f) EDS colour map showing elemental distribution of Bi, Te, Fe and S across the heterojunction, (g) Raman spectra of FeS$_2$-Bi$_2$Te$_3$ heterojunction (h) XPS survey plot of FeS$_2$-Bi$_2$Te$_3$ pellet, (i-k) Binding energy plots of Fe 2p, Bi 4f, S 2p and Te 3d.

## Electrical studies of FeS$_2$-Bi$_2$Te$_3$ heterojunction

To understand the electronic nature of the Bi$_2$Te$_3$-FeS$_2$ heterojunction, we performed voltage-current characteristics (V-I) studies. The V-I characteristics of Bi$_2$Te$_3$ and FeS$_2$ revealed an ohmic behaviour, as evidenced by the linear curve in **Figure 3a**. The ohmic nature of the Bi$_2$Te$_3$-FeS$_2$ can be attributed to the similar work function of both Bi$_2$Te$_3$ (5.1 eV) [32] and FeS$_2$ (5.2 eV) [33]. To make this pellet sensitive to Radio frequencies (RF), we created a Schottky junction at FeS$_2$ by utilising a Ti metal tip of diameter 1mm. **Figure 3b** shows the temperature-dependent Schottky diode characteristics, and **Figure 3c** illustrates the temperature-dependent forward bias characteristics of the diode, revealing a turn-on voltage of 0.18 V. To understand the diode mechanism, we calculated the Schottky barrier height of the diode using the following equation:

$$\Phi_B = \frac{kT}{q} ln\left(\frac{A^*AT^2}{I_0}\right) \quad (1)$$

Where $\Phi_B$ is the barrier height, $k$ is the Boltzmann constant, $T$ is the temperature, $q$ is the charge of electrons, $A^*$ is the Richardson constant, $A$ is the area of cross-section, and $I_0$ is the reverse saturation current obtained from **Figure 3d** [34]. The Richardson constant is calculated from the slope of **Figure 3e**. The temperature-dependent Schottky barrier height (SBH) is plotted in **Figure 3f**. A low SBH of 0.28 eV is observed between Ti and FeS$_2$, and as the temperature increases, the SBH also increases. This is due to the presence of inhomogeneities at the metal-semiconductor heterojunction that lead to Fermi-level pinning. As the temperature increases, more thermally activated electrons contribute to the current flow through thermionic emission. Consequently, the higher barrier regions become more dominant in the current flow, leading to an increase in the apparent barrier height [35–37]. For RF detector application, an ideal diode should have a low forward voltage drop and a low SBH to detect weak RF signals. The observed electrical properties of the Ti/FeS$_2$ junction satisfy the above conditions, making it an ideal diode for RF detection applications.

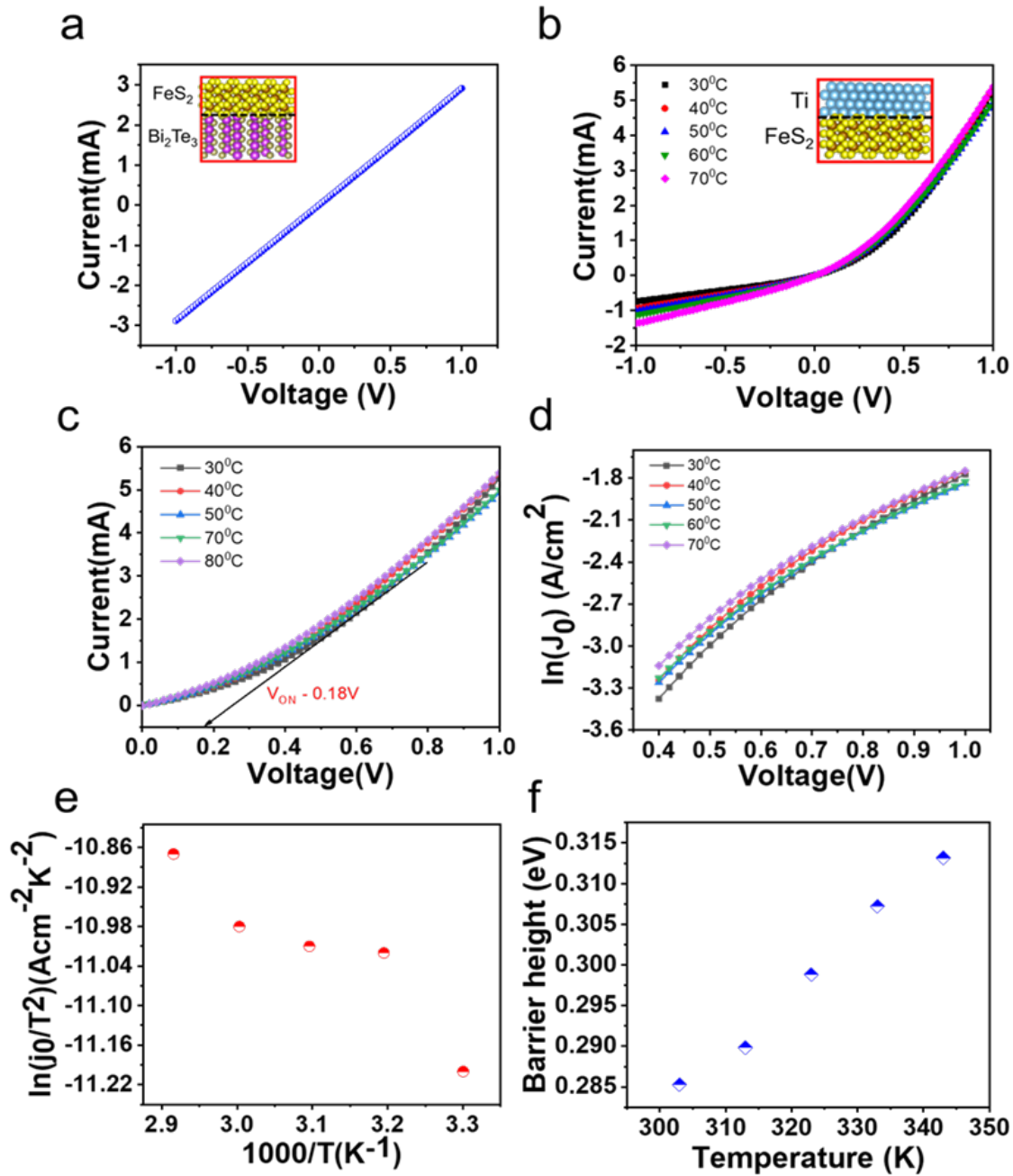

**Figure 3** - (a) Voltage-current characteristics of Bi$_2$Te$_3$-FeS$_2$, (b) Temperature-dependent Voltage-current characteristics of Ti-FeS$_2$, (c) Temperature-dependent forward voltage-current characteristics of Ti-FeS$_2$ showing turn-on voltage at 0.18V, (d) ln(j$_0$)-V plot for extracting saturation current, (e) Richardson plot, and (f) Temperature-dependent Schottky barrier height plot of Ti/FeS$_2$ junction.

# Electronic structure and thermoelectric behaviour of FeS$_2$-Bi$_2$Te$_3$ heterojunction through Density Functional theory simulations

First-principles calculations based on Density Functional Theory (DFT) were carried out to understand the electronic structure and thermoelectric properties of the FeS$_2$-Bi$_2$Te$_3$ heterojunction. The heterojunction was modelled based on experimental characterization data consisting of Bi$_2$Te$_3$ on top of the FeS$_2$ (001) surface. However, both FeS$_2$ and Bi$_2$Te$_3$ have different crystal symmetries: the former is orthorhombic while the latter is rhombohedral. Thus, to ensure commensurability, we built a model based on the modification of Bi$_2$Te$_3$ symmetry to an orthorhombic-like one. The model is shown in **Figure 4a**, beginning with a slab of FeS$_2$ (001) surface with 2 layers, we take an orthorhombic piece of Bi$_2$Te$_3$ supercell that minimizes the lattice mismatch with the FeS$_2$ (001) surface. The sublattice taken from bulk Bi$_2$Te$_3$ has the following lattice parameters: a = 7.67, b = 4.43 A. When a (3x1x1) supercell of modified Bi$_2$Te$_3$ is placed on top of a (7x1x1) supercell of FeS$_2$ (001) slab (**Figure 4a**), the lattice mismatch is minimized within 2%. Then, both structures are assembled with an initial separation (height h) of 2.5 A. Atoms on the first two layers of FeS$_2$ (001) and the top layer of Bi$_2$Te$_3$ are frozen to mimic bulk conditions, while the atoms at the heterojunction are allowed to relax according to the computational details discussed earlier. It is worth mentioning that the modified Bi$_2$Te$_3$ is not stable when considered as a free-standing structure, as our optimization calculations show a massive shrinking of the cell. On the other hand, we stress that these structures are only considered to enable a feasible computational cost on DFT simulations. To explore the interactions on the heterojunction, we performed differential charge analysis (**Figure 4b**), revealing charge depletion from the Te atoms and their rearrangement near the S atoms at the heterojunction, showcasing the significant interaction between both structures. In contrast with bulk FeS$_2$ and Bi$_2$Te$_3$, the electronic structure of the heterojunction, shown in **Figure 4c**, has partially filled valence bands, evidencing its metallic behavior. This observation agrees with electrical measurements of FeS$_2$-Bi$_2$Te$_3$ showing an ohmic behavior, indicating a metallic nature (**Figure 3a**). To gain further insights, we have projected these bands into the FeS$_2$ (001) slab and Bi$_2$Te$_3$ counterparts as represented by the colour map. It is noticeable how the states near the Fermi level are dominated by FeS$_2$ (001), which is metallic, as shown in **Figure 4c**. Interestingly, no dispersion relation is observed at the path $\Gamma \rightarrow Z$ in the reciprocal space, since all the bands are flat within the range of energies investigated. Moreover, in this direction, there is a bandgap of roughly 0.06 eV, indicating the smaller electronic conduction at the heterojunction.

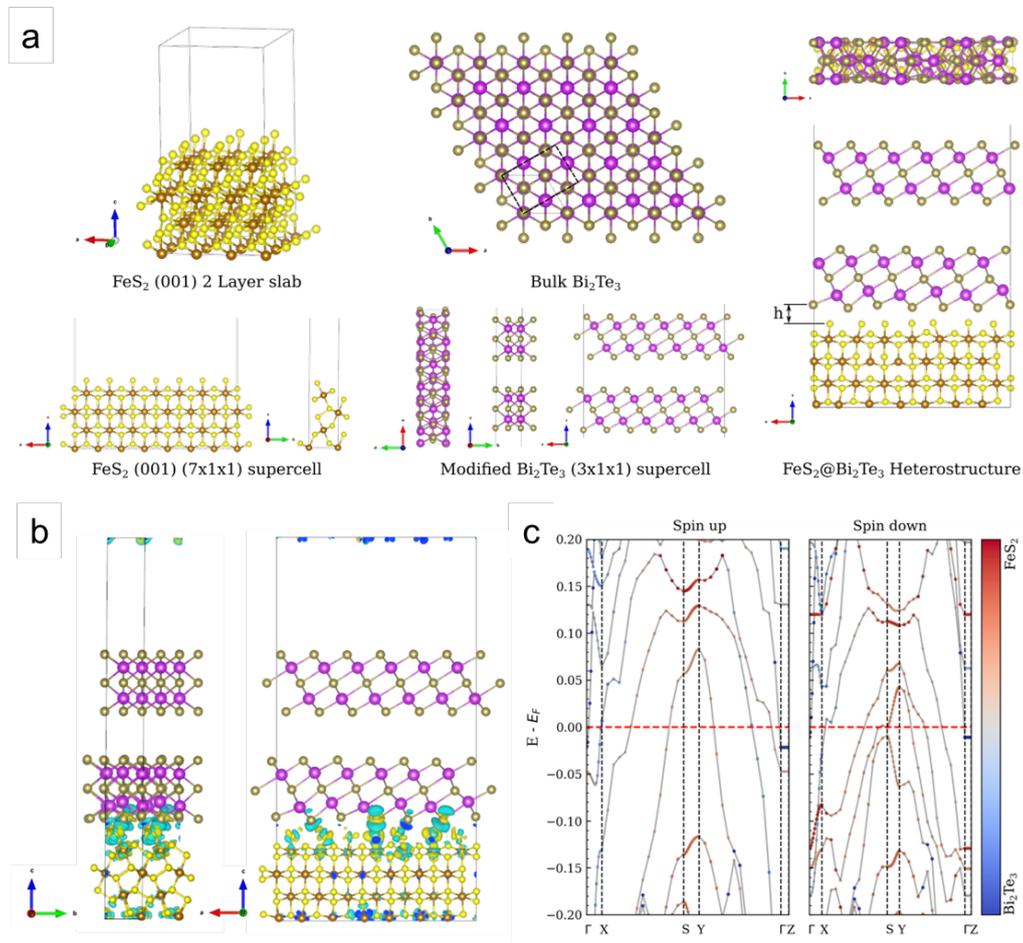

**Figure 4**: (a) The FeS$_2$-Bi$_2$Te$_3$ Heterojunction is constructed by combining 2 monolayers of a (3x1x1) supercell of modified Bi$_2$Te$_3$ (taken as a sublattice of bulk Bi$_2$Te$_3$), on top of a (7x1x1) supercell of 2L FeS$_2$ (001) slab, (b) The charge density difference analysis for the FeS$_2$-Bi$_2$Te$_3$ heterojunction. The yellow (cyan) regions represent charge accumulation (depletion).

Isosurface was set to 0.003 e/A³, (c) The electronic band structure of the heterojunction projected into the FeS₂ and Bi₂Te₃ counterparts.

We now discuss the thermoelectric properties of Bi₂Te₃, FeS₂, and FeS₂-Bi₂Te₃ heterojunction. Bulk Bi₂Te₃ crystallizes in Rhombohedral symmetry (R-3m space group). The optimized lattice parameters are a = b = 4.439 A and c = 30.661 A with $\alpha = \beta = 90°$ and $\gamma = 120°$. From the electronic perspective, bulk Bi₂Te₃ is a semiconductor with a narrow direct gap of 0.18 eV (supporting information **Figure S1.a**), by previous results [38,39]. **Figure S1.b** in the Supporting Information shows the temperature dependence of the Seebeck coefficient (S) along the x, y, and z crystallographic directions. It can be observed that the narrow band gap results in nearly vanishing Seebeck coefficients along the x and y directions. The z component shows a small value of 20 µV/K which decreases with increasing temperature. The same calculations were performed for FeS₂ (pyrite), with optimized lattice parameters of a = 3.402, b = 4.443 and c = 5.421 A with $\alpha = \beta = \gamma = 90.0°$. The obtained electronic structure shows a semiconductor with an indirect bandgap of 1.16 eV (Supporting information **Figure S2.a**), larger than previous studies that reported a value close to 0.95 eV [40,41]. Bulk FeS₂ has a high Seebeck coefficient reaching 1600 µV/K with a distinct behavior: all the components vanish below 300K with a discontinuous drop at approximately 307 K for the z component, 305 K for the y component, and 290 K for the x component. Above 315 K, all the components decrease with increasing temperature (Supporting information **Figure S2.b**). The anisotropic behavior between the x, y, and z components is attributed to the low symmetry of the orthorhombic crystal lattice.

As shown in **Figure 5a**, the y component of the Seebeck coefficient (S) is very low (5 µV/K) with linear increasing as a function of the temperature, reaching 10 µV/K at 425 K. The $S_x$ component has a negative value (of approximately -20 µV/K), indicating that electrons dominate the carrier's character [42]. The asymmetry in comparison with the x component is related to the very different lattice parameters in the plane. On the other hand, the z component is larger (in absolute value), reaching -95 µV/K at 325 K. The difference in comparison with the xy plane is attributed to the almost zero electronic conductivity ($\sigma$) in the z direction, as observed in **Figure 5b**, in which it can be seen that the conductivity decreases to zero from the y, x, and z components. Therefore, the larger Seebeck coefficient in the z direction is a combination of the zero electronic conductivity in this direction (an indication of the bandgap, in agreement with the previous discussion) and the electron concentration at the heterojunction, as shown by the charge density difference plot (**Figure 4b**). From the available data, the thermoelectric conversion capacity of the $FeS_2$–$Bi_2Te_3$ heterojunction can be evaluated by calculating the power factor (PF), expressed as $PF=S^2\sigma/\tau_0$, where S is the Seebeck coefficient, $\sigma$ is the electrical conductivity, and $\tau_0$ is the relaxation time (**Figure 5c**). Analysis reveals that the z-component of the power factor in the $FeS_2$–$Bi_2Te_3$ system is limited primarily by its relatively low electronic conductivity in that direction. In this context, maximum thermoelectric performance along the z-axis can be achieved through an optimal combination of intermediate Seebeck values and improved electronic transport. Importantly, introducing a localized heating source, as is the case in RF-induced heating of $FeS_2$, can induce a significant thermal gradient across the interface, enhancing the Seebeck voltage across the $Bi_2Te_3$ layer. This enhancement arises from the non-uniform temperature distribution at the interface, which promotes directional charge carrier diffusion and improves thermoelectric output. Thus, localized heating via RF absorption not only initiates the energy conversion process but also strategically enhances thermoelectric response by leveraging the anisotropic transport properties of the heterojunction components.

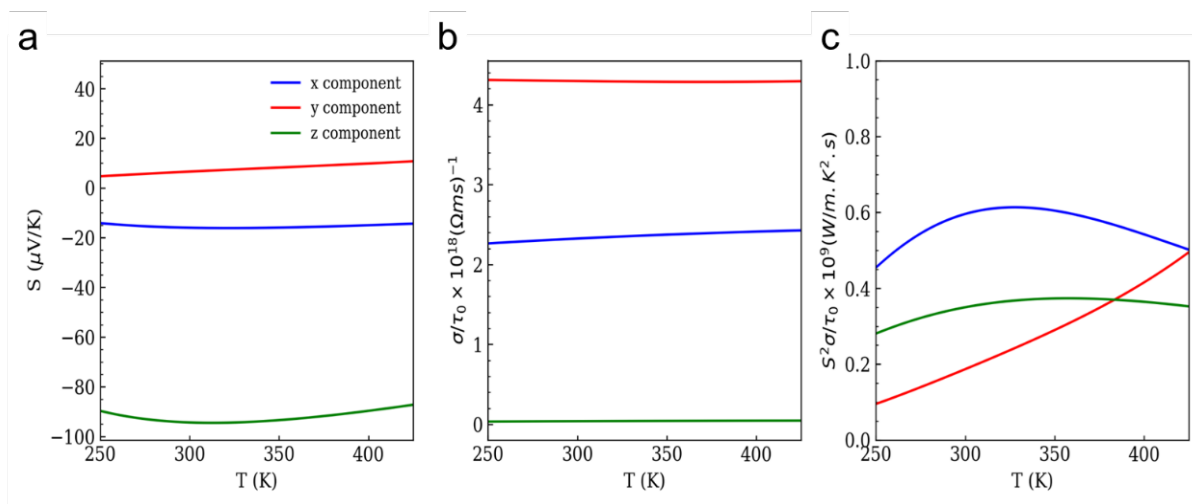

**Figure 5-**Thermoelectric properties of the FeS$_2$-Bi$_2$Te$_3$ heterojunction: (a) Seebeck coefficient components (b) The electronic conductivity components and (c) components of the Power factor ($S^2\,\sigma/\tau$) as a function of the temperature.

## RF heating studies

To investigate the localized RF heating-induced thermoelectric effect at the Bi$_2$Te$_3$–FeS$_2$ heterojunction, a series of controlled RF heating experiments were performed. **Figure 6a** presents the schematic of the electrode configuration employed during the measurements, with a corresponding experimental setup digital photograph included in the Supporting Information (**Figure S3**). A custom-built RF generator was developed for this study, capable of producing RF signals in the 35–40 MHz range with a maximum output power of 1 W. The RF signal was delivered via a 2-meter coaxial cable to a monopole antenna. The Ti–FeS$_2$ side of the heterojunction was exposed to these RF signals by placing the antenna approximately 10 cm from the device. Resulting temperature distributions were captured using thermal imaging, as shown in **Figure 6b,** and the corresponding frequency-dependent maximum temperature plot is shown in **Figure 6c**. Experimental observations indicate that the maximum recorded temperature was ~ 46 °C at 35 MHz, which exhibited a gradual decline with an increased frequency up to 38 MHz. Simultaneously, the output voltage and current across Bi$_2$Te$_3$ (**Figure**

**6d**) were measured to evaluate the power generated by RF heating. The maximum recorded voltage output from the device was 7.5 mV at 35 MHz, with a gradual decline in voltage observed at higher frequencies. This behaviour suggests that 35 MHz corresponds to the resonant frequency of the transmitting antenna system, which includes the 2-meter coaxial feed and the 75 cm monopole antenna. At resonance, efficient impedance matching and strong electromagnetic coupling occur, leading to enhanced RF absorption by the device and localized heating, which translates into a higher thermoelectric voltage. The observed peak temperature and voltage at this frequency further support this hypothesis. However, to conclusively attribute this effect to resonance, systematic experiments involving variations in antenna geometry, cable length, and impedance matching conditions are necessary. These investigations will be carried out in future studies to establish a robust understanding of the resonance-driven energy conversion mechanism. Finally, the overall power generated and power density by RF heating are shown in **Figures 6e** and **6f,** calculated by taking the product of output voltage with current and current density. The heterojunction pellet delivered a maximum power output of 100 μW and a power density of 13 mW/cm². Multiple such heterojunctions can be cascaded, and when integrated with an appropriate network matching circuit, the overall power performance can be significantly enhanced. A comparison of the output power from previously reported RF energy harvesting methods is shown in **Figure 6g**. Most studies focus on designing specific antennas and rectifiers (rectenna) to capture RF signals at various frequencies [43–47]. The collected RF signal is then fed into rectifiers to generate a DC voltage. In contrast, our work engineers the material itself to convert RF energy into heat, which is subsequently converted into electricity by a thermoelectric material. While the $FeS_2$-$Bi_2Te_3$ heterojunction demonstrates a promising approach to RF energy harvesting, its current power output of ~13 mW/cm² remains relatively low for large-scale applications. To improve efficiency, several key strategies can be explored. First, interfacial engineering between $FeS_2$ and $Bi_2Te_3$ can minimize thermal boundary resistance, allowing more efficient heat transfer and thermoelectric conversion. Incorporating high RF-absorptive materials such as graphene, carbon nanotubes (CNTs), or MXenes could enhance electromagnetic energy capture, leading to greater localized heating. Additionally, tuning the Schottky barrier at the Ti/FeS$_2$ heterojunction through work function modification or alternative metal contacts (e.g., Pt, Pd) could optimize charge transport and thermionic emission, further boosting heat generation. Improvements in thermoelectric materials such as nanostructured Bi$_2$Te$_3$, high-entropy alloys, or defect-engineered materials could also increase the figure of merit (ZT), leading to greater power output. Moreover, optimizing the RF frequency response and device geometry could enable broader spectral absorption, maximizing energy harvesting across multiple frequency bands. By implementing these strategies, the proposed RF-to-thermoelectric energy conversion approach could be further optimized for self-powered IoT sensors, wireless communication systems, and low-power electronic devices.

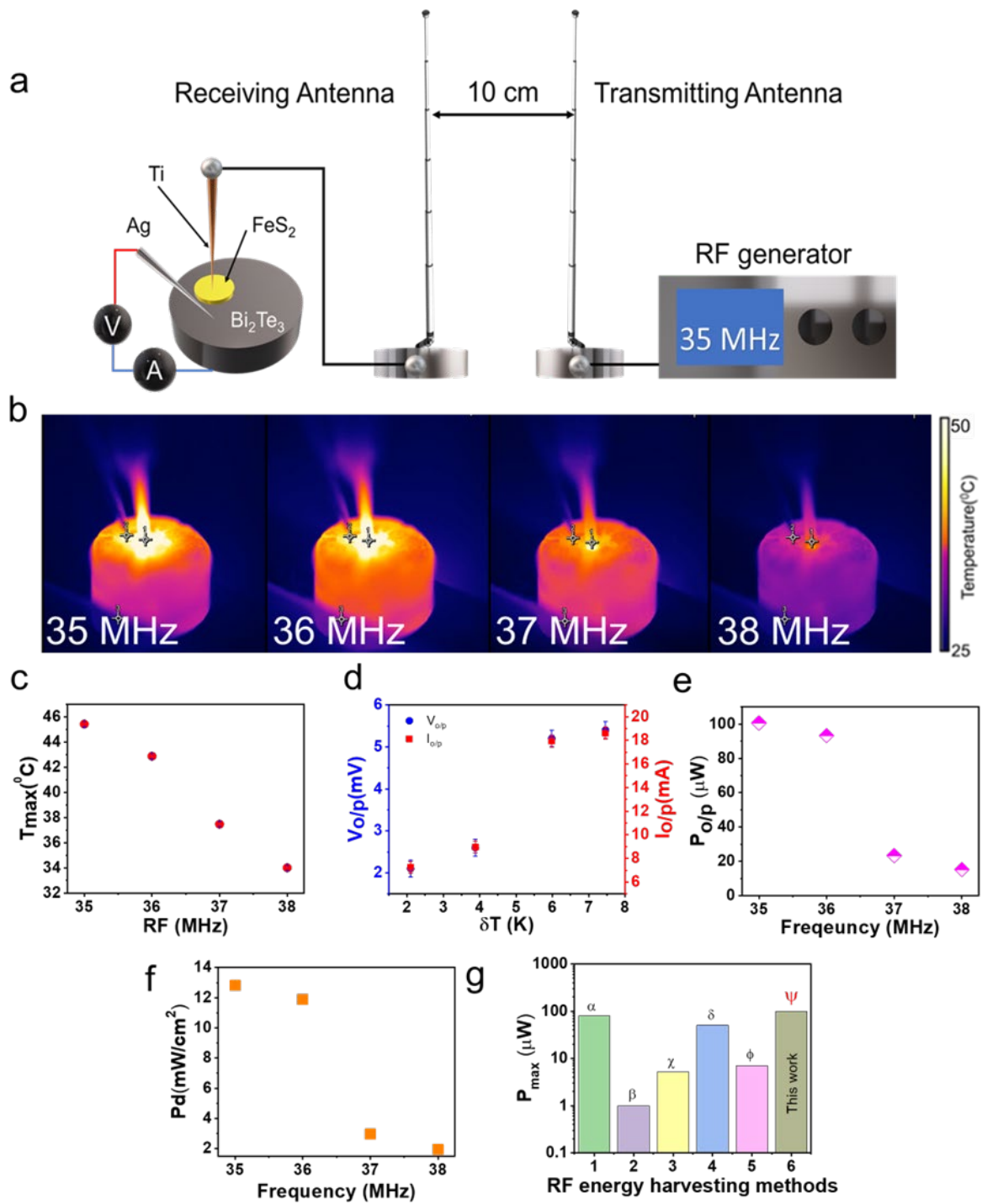

**Figure 6**- (a) Schematic representation of experimental setup for studying RF heating induced thermoelectric effect in $FeS_2$- $Bi_2Te_3$ heterojunction, (b) Thermal images RF induced heating $FeS_2$- $Bi_2Te_3$ heterojunction, (c) Frequency-dependent maximum temperature plots, (d) Frequency-dependent output voltage and current plots, (e) Frequency-dependent power plots,(f) Frequency dependent power density plots, (g) Comparison power plots of other RF energy harvesting approaches (α- Receiving antenna Wi-Fi, GSM 900,1800 MHz, β-Textile antenna Wi-Fi, γ-Patch antenna Wi-Fi, GSM, δ- Multi resonator antenna 915 MHz, Φ-Rectenna Wi-Fi, GSM, Ψ-RF-TE composite HF 35-38 MHz this work)

## Molecular Dynamics Calculations

To explain the mechanisms of thermal transport in RF-heated $Bi_2Te_3$, we performed a series of molecular dynamics calculations. **Figure 7a** shows the structure of $Bi_2Te_3$ used for MD simulations. The thermal conductivities of $Bi_2Te_3$ in three directions calculated by MD simulations are presented in **Figure 7b**. The thermal conductivity generally decreases within a temperature range of 300 to 425 K. This can be attributed to multiple factors, such as defects (e.g., grain boundaries and vacancies) in $Bi_2Te_3$ crystal at higher temperatures; an additional reason could be heterogeneities at bulk scale as compared to an ideal system considered for MD simulations. Interestingly, different trends of thermal conductivity as a function of temperature are reported in the literature. Huang and Kaviany [48] show that the thermal conductivity of $Bi_2Te_3$ decreases with temperature using MD simulations, which agrees with the experimental results by Satterthwaite and Ure [49]. However, Zhang et al. [50] report an increase in thermal conductivity of nanostructured $Bi_2Te_3$ with temperature. Our MD simulations also reveal the anisotropy in $Bi_2Te_3$, where the thermal conductivity is higher in the x and y directions (in-plane directions) than z direction (cross-plane direction) due to different governing interactions between atoms. The same anisotropic behaviour is reported in previous studies [51–53]. **Figure 7(d-f)** presents the spatial temperature distributions of the simulation box at different stages of simulated RF heating. Temperature is evenly distributed at the early stage (0 to 0.1 ns) since the heating rate is low and little thermal energy is added to the system (**Figure 7d**). When the system is heated at a higher rate, thermal energy starts to accumulate in the heat source and propagates through the system, creating a significant thermal gradient (**Figure 7e**). At 0.9 to 1 ns, the gradient becomes more pronounced and a temperature difference up to 200 K is observed due to the low thermal conductivity of $Bi_2Te_3$ in the z-direction (**Figure 7f**).

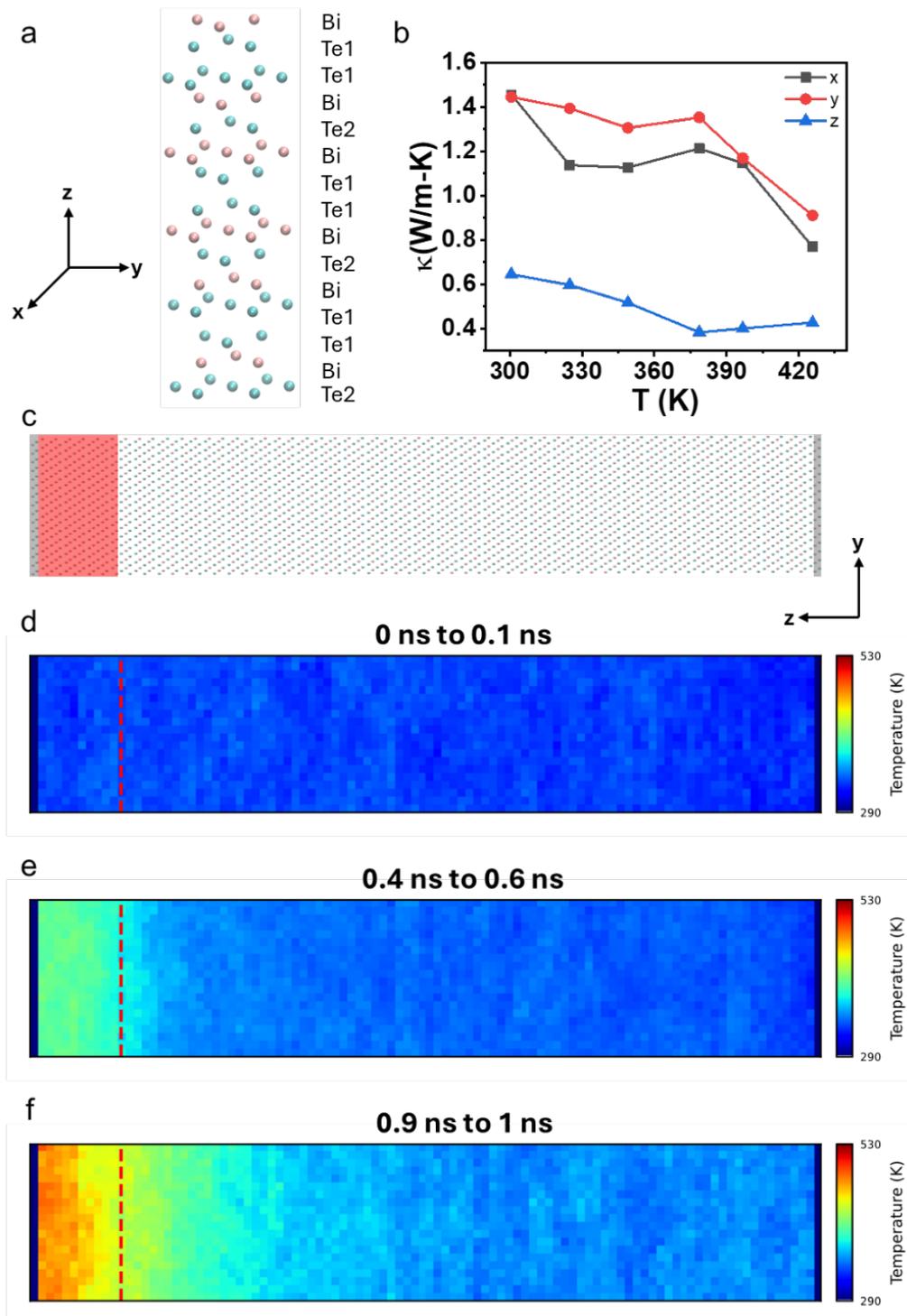

**Figure 7**- Molecular dynamics simulations of $Bi_2Te_3$. (a) The hexagonal conventional cell of $Bi_2Te_3$, (b) Calculated thermal conductivities of $Bi_2Te_3$ in three directions as a function of temperature, (c) NEMD setup to simulate RF heating where the heat source is the region coloured in red, (d-f) The temperature distributions in the system at different stages during RF heating.

With the thermal gradient induced by RF heating in $Bi_2Te_3$ confirmed through molecular dynamics (MD) simulations, the RF-induced thermoelectric effect in $FeS_2$-$Bi_2Te_3$ is further discussed below. When an RF signal is applied to the metal-semiconductor junction, in this case, the Ti-$FeS_2$ heterojunction, the incident electromagnetic energy interacts with the charge carriers at the junction. These carriers gain sufficient energy from the RF field to overcome the Schottky barrier at the heterojunction, allowing for the flow of charge across the junction. This process, known as the Schottky barrier lowering, results in an increased carrier concentration and the excitation of electrons in the metal (Ti) and semiconductor ($FeS_2$) regions[54]. The RF energy also generates localized heating at the Ti–$FeS_2$ junction due to the interaction of the RF signal with the material's free electrons, which causes them to vibrate and collide with atoms, thus raising the temperature in this localized region. This localized heating leads to the establishment of a thermal gradient across the heterojunction. $Bi_2Te_3$, which is near this heated junction, benefits from this thermal gradient. Due to the thermoelectric properties of $Bi_2Te_3$, this temperature difference causes charge carriers in the material to diffuse from the hot region to the cooler region, creating a voltage gradient across the $Bi_2Te_3$ layer. The above concept is depicted schematically as shown in **Figure 8**. The generated thermoelectric voltage is a direct result of the Seebeck effect, wherein the heat energy is converted into electrical energy. In this case, the localized heat at the Ti–$FeS_2$ heterojunction essentially drives the thermoelectric effect in $Bi_2Te_3$, thereby enabling the conversion of the thermal energy generated by the RF signal into usable electrical power. This localized heating not only enhances the efficiency of the thermoelectric effect but also demonstrates a unique coupling between RF energy absorption and thermoelectric energy harvesting.

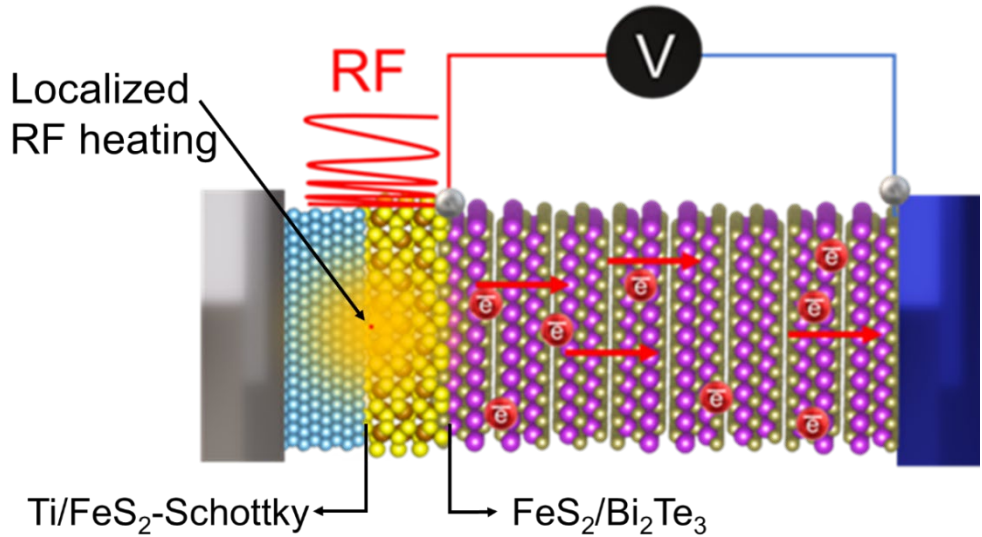

**Figure 8-** Schematic representation of localised RF heating induced thermoelectric effect in FeS₂-Bi₂Te₃ heterojunction.

## Conclusions

In this study, we have demonstrated a FeS$_2$-Bi$_2$Te$_3$ heterojunction-based RF thermoelectric energy harvesting device, where a Schottky junction at the Ti/FeS$_2$ heterojunction enables the conversion of RF energy into localized heating, which is subsequently transformed into electricity using Bi$_2$Te$_3$. The RF heating studies revealed a peak temperature of 45.5 $^0$C at 35 MHz with 1W RF power, highlighting the efficient absorption of electromagnetic energy. Under a thermal gradient of 5.5 K, the FeS$_2$-Bi$_2$Te$_3$ heterojunction achieved a maximum power output of 100μW, demonstrating its capability for energy harvesting. Density functional theory (DFT) and molecular dynamics (MD) simulations provided insights into the interfacial charge and heat transport mechanisms, confirming anisotropic thermoelectric properties and electron accumulation at the heterojunction. These findings establish a new paradigm for RF to thermal electric energy conversion, offering an alternative approach to conventional rectenna-based RF energy harvesting. Future work will focus on optimizing interfacial engineering, integrating nanostructured thermoelectric material, and enhancing RF absorption efficiency for real-world applications in self-powered IoT devices and wireless sensors.

# Experimental Methodologies

## Fabrication of $FeS_2$-$Bi_2Te_3$ heterojunction

3 g of $Bi_2Te_3$ powder was first placed into a 10 mm diameter die and cold-pressed. Subsequently, 2 g of bulk $FeS_2$ pieces were added on top, and the composite was cold-pressed under a load of 1 ton for 5 minutes. This process yielded a $FeS_2$–$Bi_2Te_3$ heterojunction pellet.

## Characterization studies

The crystal structure and Phase analysis of $FeS_2$-$Bi_2Te_3$ were performed using X-ray diffraction (XRD) (Bruker D8 Advance with a Lynx eye detector using Cu-Kα radiation). The elemental composition and morphology analysis were done through Scanning electron microscopy (SEM) and energy dispersive spectroscopic studies (EDS) (ZEISS Sigma). The vibrational nature of the $FeS_2$-$Bi_2Te_3$ heterojunction was also analyzed using Raman Spectroscopy (WITec, UHTS 300VIS, Germany). The chemical oxidation states were analyzed using X-ray Photoelectron Spectroscopy (XPS) (Thermo Fisher Scientific makes Nexsa base). The voltage-current characteristics and dielectric measurements of Ti-$FeS_2$-$Bi_2Te_3$ were done using a Source meter (Keithley 2450) and a Precision LCR meter (SM6026). The thermoelectric measurements were done in the Physical Property Measurement System (PPMS). RF heating and thermal imaging studies were done using a lab-made RF signal generator of 1W power and a Thermal imaging camera (Optris PI640). Thermal voltage measurements during RF heating were done Moku voltage measurement instrument (Moku: Go demo).

# Computational Details

## Ab initio details

Spin-polarized ab initio calculations based on Density Functional Theory (DFT) were carried out using the Vienna Ab-initio Simulation Package (VASP) [55] with the Projected Augmented Wave pseudopotential method [56]. The valence configuration adopted for the atoms in the system was: 3d7 4s1 for Fe, 3s2 3p4 for S, 6s2 6p3 for Bi, and 5s2 4p4 for Te. Khon-Sham orbitals were expanded using a plane-wave basis set with a kinetic energy cutoff of 300 eV. The exchange and correlation interactions were approximated by the revised non-local Vydrov and Van Vooris (rVV10) functional as implemented by Sabatini et al. [57]. The DFT+U approach [58] was applied to describe the localization of d electrons in Fe, setting the U parameter to 2 eV in accordance with previous calculations [59]. The Brillouin zone was sampled by a uniform grid of k-points within the scheme of Monk Horst and Pack [60]; the specific k-point mesh is specified in the following discussions for each structure. Self-consistency threshold was set to $10^{-7}$ eV, and forces were minimized below $10^{-2}$ eV/A. Data analysis for electronic structure calculations was carried out with the VASPKIT suite [61]. Thermoelectric properties of $FeS_2$-$Bi_2Te_3$ heterojunction were calculated within the semiclassical Boltzmann Theory as implemented in the Boltz Trap software [62]. The Seebeck coefficient (S), electronic conductivity ($\sigma$) and the electronic contribution to the lattice thermal conductivity ($\kappa_e$) are calculated from the KS eigenvalues which are used to interpolate analytic expressions as a function of the chemical potential ($\mu$) temperature (T) and relaxation time ($\tau_0$).

**Molecular Dynamics details**

Molecular dynamics (MD) simulations are performed to gain insights into thermal transport in $Bi_2Te_3$. We only simulate $Bi_2Te_3$ instead of $FeS_2$/$Bi_2Te_3$ heterojunction due to differences in lattice structures and absence of appropriate cross-interaction parameters in the literature. $Bi_2Te_3$ has a rhombohedral lattice structure with a hexagonal conventional cell with three quintuple layers. We define two orthogonal directions in the hexagonal plane as x and y directions (i.e., in-plane directions) and the cross-plane direction as z direction, as shown in **Figure 7a.** There are two types of Te atoms with two bonding environments. Te1 atoms have weaker interlayer van der Waals interactions with other atoms, while Te2 atoms form strong intralayer bonds. The short-range interactions within $Bi_2Te_3$ are described in the form of a Morse potential developed by Qiu and Ruan [63]

$$\varphi_s(r_{ij}) = D_e\{[1 - e^{-a(r_{ij}-r_0)}]^2 - 1\},$$

where $\varphi_s$ is the interatomic potential, $r_{ij}$ is the distance between atoms $i$ and $j$, and $D_e, a, r_0$ are parameters whose values can be found in Ref [64]. The long-range Coulombic interactions are treated by the particle-particle particle-mesh (PPPM) method. The MD simulations are carried out using the Large-scale Atomic/Molecular Massively Parallel Simulator (LAMMPS) package [65].

A cubic simulation box with a dimension of $60 \times 60 \times 60$ Å$^3$ It is created to calculate thermal conductivity. The system is minimized and equilibrated in the NPT and NVT ensembles for 2 ns. The temperature and pressure damping parameters are 0.1 and 1 ps, respectively. The lattice constants obtained from the equilibrated simulation box at 300 K are $a = 4.343$ Å and $c = 30.443$ Å which agree well with previous experimental results [66]. We carry out a production run for 20 ns where heat fluxes are recorded in x, y, and z directions. We calculate the thermal conductivity using the Green-Kubo method [67] by integrating the heat flux autocorrelation function:

$$\kappa = \frac{V}{k_B T^2} \int_0^\infty \langle J(t_0)J(t_0+t)\rangle\, dt$$

where $V$ is volume, $k_B$ is the Boltzmann constant, $T$ is temperature, $J$ is heat flux, $t$ is the correlation time, and $\langle \rangle$ is the average overall time of origins $t_0$. The production period is divided into 20 trajectories of 1 ns, and the thermal conductivity is averaged over 20 trajectories.

Nonequilibrium molecular dynamics (NEMD) simulations are carried out to simulate the heat transfer in the cross-plane direction under RF heating. We create a slab-shaped simulation box with dimensions of 60 Å, 60 Å and 600 Å in x, y, and z directions, respectively. The simulation box is divided into 100 bins in the z direction with equal width, and the leftmost and rightmost bins are fixed as walls. Periodic boundary conditions are applied in the x and y directions only. We consider the 10 bins nearest to the left wall an RF heat source (**Figure 7c**). Since a continuous function of variable heating rate cannot be defined in LAMMPS, a stepwise heating rate ranging from 0.0016 to 0.01 kcal/mol/fs is applied to the heat source for 1 ns to emulate the sinusoidal heating by RF (supporting information **Figure S4**). The spatial temperature distribution of the simulation box is recorded during the heating process.